\pdfoutput=1
\documentclass[conference]{IEEEtran}
\IEEEoverridecommandlockouts
\usepackage{cite}
\usepackage{amsmath,amssymb,amsfonts}
\usepackage{algorithmic}
\usepackage{graphicx}
\usepackage{textcomp}
\usepackage{xcolor}
    
\begin{document}

\title{Coordinated Scheduling of Electric Vehicles Within Zero Carbon Emission Hybrid AC/DC Microgrids}

\author{
\IEEEauthorblockN{1\textsuperscript{st} Reza Bayani}
\IEEEauthorblockA{\textit{ Electrical and Computer Engineering} \\
\textit{University of California, San Diego}\\
San Diego, USA 92161 \\
rbayani@ucsd.edu}
\and
\IEEEauthorblockN{2\textsuperscript{nd} Arash Farokhi Soofi}
\IEEEauthorblockA{\textit{  Electrical and Computer Engineering} \\
\textit{University of California, San Diego}\\
San Diego, USA 92161  \\
afarokhi@ucsd.edu}
\and
\IEEEauthorblockN{3\textsuperscript{nd} Saeed D. Manshadi}
\IEEEauthorblockA{\textit{ Electrical and Computer Engineering} \\
\textit{San Diego State University}\\
San Diego, USA 92182 \\
smanshadi@sdsu.edu}}

\maketitle

\begin{abstract}
    Microgrids with AC/DC architecture benefit from advantages of both AC and DC power. In this paper, daily operation problem for a zero-carbon AC/DC microgrid in presence of electric vehicles (EVs) is considered. In this framework, EVs' batteries are mobile energy storage systems, which allow desirable operation of the microgrid during peak demand hours. This study shows in absence of storage system, EVs' batteries can be properly managed to satisfy the system requirements. In the case studies, several sensitivity analyses based on variations in battery degradation costs, solar irradiance, and inverter capacity are investigated.
\end{abstract}


\section*{Nomenclature}
\subsection*{Variables}
\noindent \begin{tabular}{ l p{6.55cm} }
$E$ & Energy level of EV\\
$I$ & Binary variable\\
$p,q,s$ & Real/Reactive/Apparent power\\
$v,\theta$ & Voltage magnitude and angle\\
\end{tabular}

\subsection*{Sets}
\noindent \begin{tabular}{ l p{6.55cm} }
$\mathcal{C}_i$ & Set of Inverters placed on bus i\\
$\mathcal{E}_i$ & Set of EVs placed on bus i\\
$\mathcal{L}_i$ & Set of loads placed on bus i\\
$\mathcal{S}_i$ & Set of PV units placed on bus i\\
$\mathcal{W}_i$ & Set of WT units placed on bus i\\
\end{tabular}

\subsection*{Indices}
\noindent \begin{tabular}{ l p{6.55cm} }
$c$ & Inverter\\
$ch,dis,v2g$ & Charge/Discharge/V2G power of EV\\
$d$ & Demand served\\
$D$ & Requested demand\\
$e$ & EV number\\
$F$ & Forecast output of renewable unit\\
$i,j,o$ & Bus number\\
$s,w$ & Solar/Wind\\
$t$ & time\\
$to,fr$ & To/From\\
\end{tabular}

\subsection*{Parameters}
\noindent \begin{tabular}{ l p{6.55cm} }
$G,B$ & Real/Imaginary part of admittance matrix\\
$K_{L}$ & Value of lost load\\
$O$ & Parameter for modeling EV travel status\\
$r$ & Resistance of line\\
$\beta_1,\beta_2$ & Degradation cost multipliers\\
$\xi$ & Coefficient for estimating apparent power\\
\end{tabular}

\section{Introduction}
In order to enhance reliability and resilience of power systems and mitigate environmental concerns, microgrids are introduced \cite{abazari2019coordination}. Microgrids are composed of distributed energy resources as well storage units, which allows them to be operated independent of the main grid\cite{ghasemi2020multi,rahmanzadeh2018optimal,shojaeighadikolaei2020demand}. This means that a microgrid is able to generate enough power to fulfil its demand, if operational constraints are satisfied \cite{parhizi2015state}. Conventionally, electricity is delivered in the AC form. This is due to the fact that power is originally generated in the AC form, and existing power transmission and distribution infrastructure is designed to work with the AC power. However, DC power has some benefits such as smaller loss, a higher level of reliability, and less technical challenges in terms of frequency and voltage regulation \cite{Khodayar2016TheArchitecture}. In this work, a hybrid AC/DC microgrid is considered to benefit from DC perks \cite{manshadi2016decentralized,bayani2020short}.
\par The transportation system is experiencing a rapid shift towards electrification, and EVs are increasingly being introduced to it. One aspect of EVs is their ability to store energy. With proper implementations, EVs can inject power to the grid as well. In our previous work, we have shown that EV fleet storage capabilities can be utilized to offer several services to the grid\cite{bayani2020autonomous}. The concept of vehicle to grid (V2G) takes advantage of EV battery, and allows Plug-in EVs (PEVs) to be utilized for power system operation \cite{ahmadian2020review,babaei2021data}. System operator can come up with stimulative plans to encourage PEV owners to participate in management schemes. In addition, researchers are investigating EV parking lots as storage systems to offer flexibility \cite{lekvan2021robust,ansari2019framework}.
\par In this work, a self-perpetuating zero-carbon microgrid is considered with sufficient amount of renewable generation to ensure reliable operation of the system. Instead of bulk energy storage systems, distributed storage in form of EV charging stations at various nodes of the system are considered. It is shown that with proper management of EV batteries, they can store adequate energy to meet owners' demand for daily trips, as well as securing reliable operation of the microgrid during insufficient renewable generation periods. The contributions of this work can be summarized as follows:
\begin{enumerate}
    \item A highly renewable integrated hybrid carbon free AC/DC microgrid with no bulk storage is introduced which operates independent of the main grid. It is shown the collective storage capacity of PEVs in the microgrid can be utilized in order to both provide energy for the owners' daily trip and the microgrid's electricity demand. This is acquired by exploiting
    the fast-charging capability of EVs through designation of charging stations to only DC side of the microgrid.
    \item The impacts of variations in V2G costs, daily solar irradiance, and inverter capacity are investigated and their implications on the operation of the microgrid are discussed in the results section.
\end{enumerate}

\section{Problem Formulation}
The set of equations governing the operation of the proposed EV integrated AC/DC microgrid structure can be categorized into several groups, which are discussed here.

\subsection{Objective Function}

The objective function is defined by \eqref{eq:gas_objective}, which is composed of two elements. The first term in objective function is the penalty for value of lost load, which is acquired through multiplying the total value of daily demand which is not served by penalty factor, $K_L$. The second term in the objective function represents the degradation cost of EV batteries as a result of V2G power injections. It is also assumed that operation costs of renewable units are negligible.\\
\begin{equation}
     \min K_{L}\cdot\sum_{t}\sum_{i}(p_{i,t}^{D}-p_{i,t}^{d}) +\sum_{t}\sum_{e}(\beta_2\cdot {p^{v2g}_{e,t}}^2 + \beta_1 \cdot p^{v2g}_{e,t})
\label{eq:gas_objective}
\end{equation}

\subsection{Operational Limitation Constraints}
To satisfy the operation requirements of the system, according to \eqref{eq:voltage_limit}, voltage magnitude of each bus should be within certain limits at all times. The real/reactive power transmitted by each inverter is also limited by \eqref{eq:inverter_limit_active} and \eqref{eq:inverter_limit_reactive}, respectively. The apparent power in AC side is acquired through the linear approximation shown in \eqref{eq:apparent_ac}. The apparent power of each line in AC side, and the real power flowing through lines in DC side are also limited by the maximum capacity of that line as shown in \eqref{eq:apparent_limit_ac} and \eqref{eq:apparent_limit_dc}, respectively.
\begin{subequations} \label{eq:ac_constraints}
\begin{alignat}{3}
& \underline{v} \leq v_{i,t} \leq \overline{v} \label{eq:voltage_limit}\\
& \underline{p_c}\leq p_{c,t}\leq \overline{p_c}\label{eq:inverter_limit_active}\\
& \underline{q_c}\leq q_{c,t}\leq \overline{q_c}\label{eq:inverter_limit_reactive}\\
& s^{ac}_{l,t} =  p^{ac}_{l,t} + \xi\cdot q^{ac}_{l,t} \label{eq:apparent_ac}\\
& \underline{s_l} \leq s^{ac}_{l,t} \leq \overline{s_l}\label{eq:apparent_limit_ac}\\
& \underline{s_l} \leq p^{dc}_{l,t} \leq \overline{s_l}\label{eq:apparent_limit_dc}
\end{alignat}
\end{subequations}

\subsection{Renewable Generation Constraints}
As defined in \eqref{eq:wt_dispatch} and \eqref{eq:pv_dispatch}, the real power dispatch of renewable resources is considered as a variable, varying between zero and the maximum available generation output of each supplier. This allows for curtailments to take place when necessary, due to operational limitations. It is also considered that each renewable supplier is equipped with an inverter that can generate reactive power in both positive and negative regions, as seen in \eqref{eq:wt_q} and \eqref{eq:pv_q}.
\begin{subequations} \label{eq:renewables}
\begin{alignat}{3}
& 0 \leq p_{w,t} \leq p^F_{w,t} \label{eq:wt_dispatch}\\
& 0 \leq p_{s,t} \leq p^F_{s,t} \label{eq:pv_dispatch}\\
& - p_{w,t} \leq q_{w,t} \leq p_{w,t} \label{eq:wt_q}\\
& -p_{s,t} \leq q_{s,t} \leq p_{s,t} \label{eq:pv_q}
\end{alignat}
\end{subequations}

\subsection{Power Flow Constraints}
To calculate the real and reactive power flowing through each line, the first-order approximate AC power flow equation shown in \eqref{eq:ac_line_active} and \eqref{eq:ac_line_reactive} is applied throughout the AC side of the system. For the DC side of the microgrid, according to \eqref{eq:dc_line_active}, the power transmitted at each line is calculated based on the line's resistance and the voltage difference of neighboring buses.
\begin{subequations} \label{eq:powerflow}
\begin{alignat}{3}
& p^{ac}_{l,t}=-G_{j,o}(v_{j,t}-v_{o,t})+B_{j,o}(\theta_{j,t}-\theta _{o,t})
\label{eq:ac_line_active}\\
& q^{ac}_{l,t}=B_{j,o}(v_{j,t}-v_{o,t})+G_{j,o}(\theta_{j,t}-\theta _{o,t})\label{eq:ac_line_reactive}\\
& p^{dc}_{l,t}=\frac{v_{j,t}-v_{o,t}}{r_l}\label{eq:dc_line_active}
\end{alignat}
\end{subequations}

\subsection{Nodal Balance Constraints}
At each bus of the system, the summation of net transmitted power through connecting lines, output of suppliers, and net injected power of EVs (discharged power in form of V2G minus charged power), should be equal to the demand served on that bus. The real power balance equation defined in \eqref{eq:ac_balance_active} holds for both AC/DC sides, whereas the nodal reactive balance displayed in \eqref{eq:ac_balance_reactive} is only applied in AC part of the microgrid.
\begin{subequations} \label{eq:powerflow}
\begin{alignat}{3}
&\begin{aligned}
  \sum_{e\in \mathcal{E}_{i}} (p^{v2g}_{e,t}-p^{ch}_{e,t})
+ \sum_{w\in \mathcal{W}_{i}} p_{w,t} 
+ \sum_{s \in \mathcal{S}_{i}} p_{s,t}\\
+ \sum_{c\in \mathcal{C}_{i}} p_{c,t} 
+ \sum_{l\in \mathcal{L}^{to}_{i}} p_{l,t}
- \sum_{l\in \mathcal{L}^{fr}_{i}} p_{l,t}
=
{p^d_{i,t}}
\label{eq:ac_balance_active}\end{aligned}\\
&\begin{aligned}
 \sum_{w\in \mathcal{W}_{i}} q_{w,t} 
+ \sum_{s \in \mathcal{S}_{i}} q_{s,t}
+ \sum_{c\in \mathcal{C}_{i}} q_{c,t} \\
+ \sum_{l\in \mathcal{L}^{to}_{i}} q_{l,t}
- \sum_{l\in \mathcal{L}^{fr}_{i}} q_{l,t}
=
{q^d_{i,t}}
\label{eq:ac_balance_reactive}\end{aligned}
\end{alignat}
\end{subequations}

\subsection{EV Model Constraints}
The EVs are either operating in one of these three modes throughout the day: traveling, plugged-in, or idle. EVs lose their stored energy when they are in the traveling mode (modeled with the parameter $O^{dis}_{t}$). If an EV is in idle mode, its energy level remains the same. Plugged-in EVs (modeled with binary parameter $O^{plug}_{t}$) can take two decisions: they either will be charged or will inject power to the grid (V2G mode), which are modeled with $I^{ch}_{e,t}$ and $I^{v2g}_{e,t}$ binary variables, respectively. Charging and V2G power of plugged EVs is modeled by \eqref{eq:ev_charge} and \eqref{eq:ev_v2g}, respectively. The energy stored in each EV's battery at each time step is calculated according to \eqref{eq:ev_energy}, and must remain within certain limits as defined by \eqref{eq:ev_energy_limit}.
\vspace{-.35cm}
\begin{subequations} \label{eq:dc_constraints}
\begin{align}
& 0 \leq p^{ch}_{e,t}\leq I^{ch}_{e,t}\cdot O^{plug}_{e,t}\cdot \overline{p_{e}} \label{eq:ev_charge}\\
& 0 \leq p^{v2g}_{e,t}\leq I^{v2g}_{e,t}\cdot O^{plug}_{e,t}\cdot \overline{p_{e}} \label{eq:ev_v2g}\\
& p^{dis}_{e,t} = O^{dis}_{e,t}\cdot\overline{p_e}\\
& I^{ch}_{e,t} + I^{v2g}_{e,t} \leq 1 \label{eq:ev_binary}\\
& E_{e,t}= E_{e,t-1}- p^{dis}_{e,t}- \frac{p^{v2g}_{e,t}}{\eta^{v2g}_{e}}+p^{ch}_{e,t}\cdot\eta^{ch}_{e}  \label{eq:ev_energy} \\
& \underline{E_e}\leq E_{e,t}\leq \overline{E_e} \label{eq:ev_energy_limit}
\end{align}
\end{subequations}
The discussed formulations in (1)-(6) together form a mixed integer quadratic programming problem. Although there is no guarantee to reach global optimal in these problems, available solvers, such as Gurobi \cite{gurobi} which was used in this research, can solve them to acceptable levels of optimality gap.

\section{Test System Configuration}
\begin{figure}[hb!]
    \centering
    \includegraphics[width=.75\linewidth]{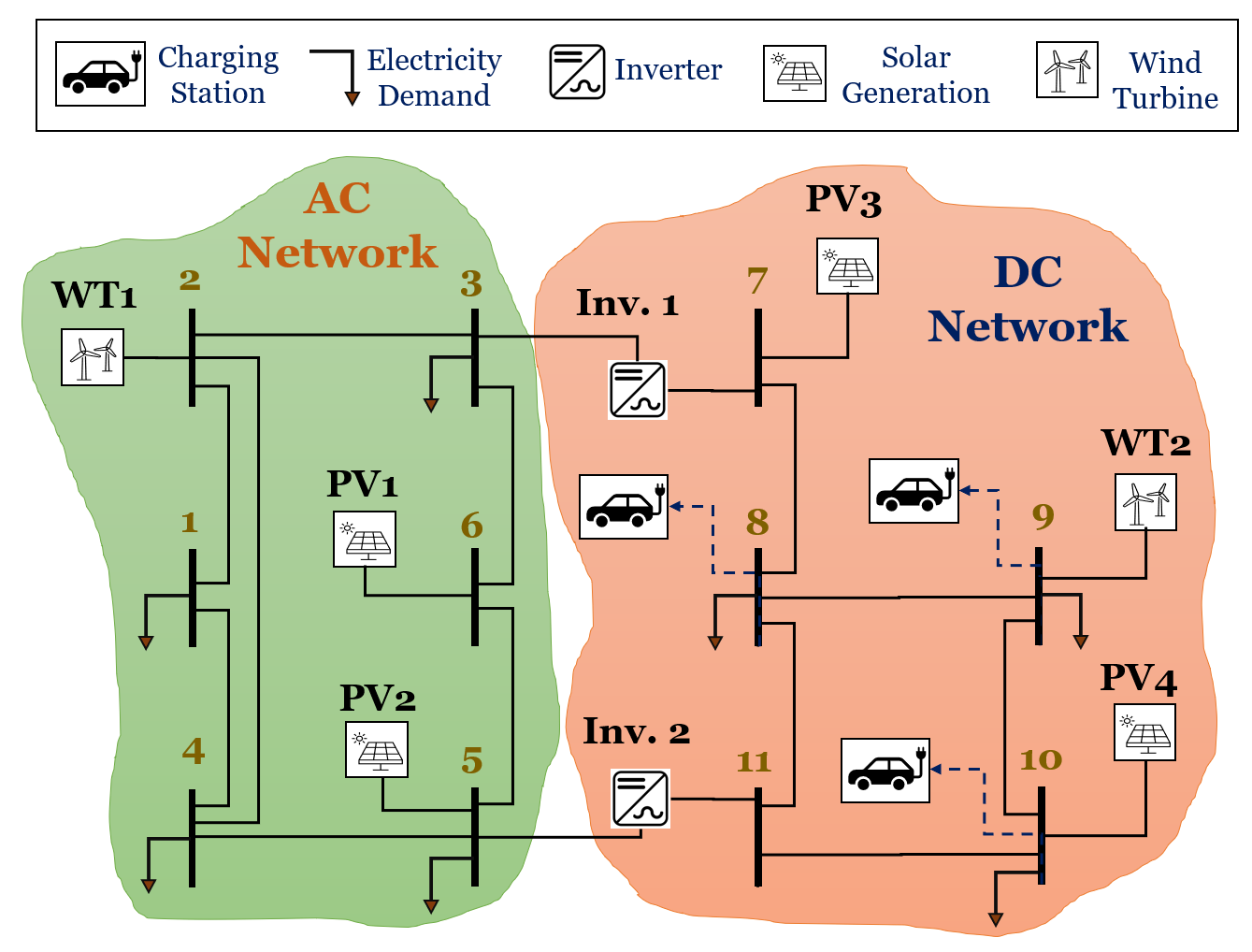}
         \vspace{-.35cm}
   \caption{Hybrid AC/DC Microgrid Structure}
    \label{fig:ac_dc_microgrid}
    \vspace{.45cm}
    \includegraphics[width=.8\linewidth]{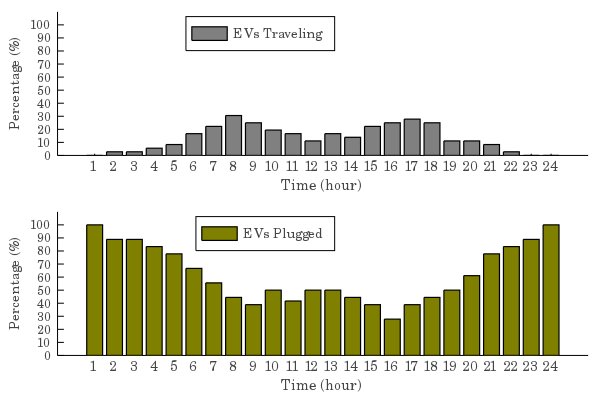}
        \vspace{-.35cm}
    \caption{Daily Distribution of Traveling and Plugged EVs}
    \label{fig:ev_flag}
    \vspace{-.55cm}
\end{figure}

The configuration of the hybrid AC/DC microgrid with high penetration level of renewable suppliers and EVs is displayed in Fig. \ref{fig:ac_dc_microgrid}. The power between AC and DC sides is exchanged through two inverters. This microgrid has 11 buses and 12 lines, with 3 charging stations which are all placed at DC side. This setup increases system efficiency by means of reducing AC to DC conversion losses for EVs. At each side, two Photovoltaic (PV) units and one Wind Turbine (WT) are placed. During the daylight hours, PV serves as the main contributor of the system's demand, whereas during early morning and later hours of the day, WT units produce the required energy. The percentage of EVs that are traveling or connected at each time of the day are displayed in Fig. \ref{fig:ev_flag}.
\section{Results and discussion}
In this section, the simulation results for the daily operation of the proposed microgrid structure are discussed under several case studies. In case 1, the grid's operation during normal conditions is illustrated. In case 2, the impact of V2G cost on the microgrid's operation is investigated. In case 3, the grid's operation under lower solar irradiances scenarios such as rainy or cloudy weather conditions is explored. Finally in case 4, the consequences of reduction in inverters' capacity are analyzed. This situation could arise as a result of failure in inverters. The penalty for lost load is considered equal to $\$1000/MWh$, and $\beta_1,\;\beta_2$ are respectively \textcent$0.1/kWh^2$ and \textcent$10/kWh$.

\subsection{Case 1 - Normal operating conditions}
\begin{figure}[b!]
    \centering
    \vspace{-.45cm}
    \includegraphics[width=.82\linewidth]{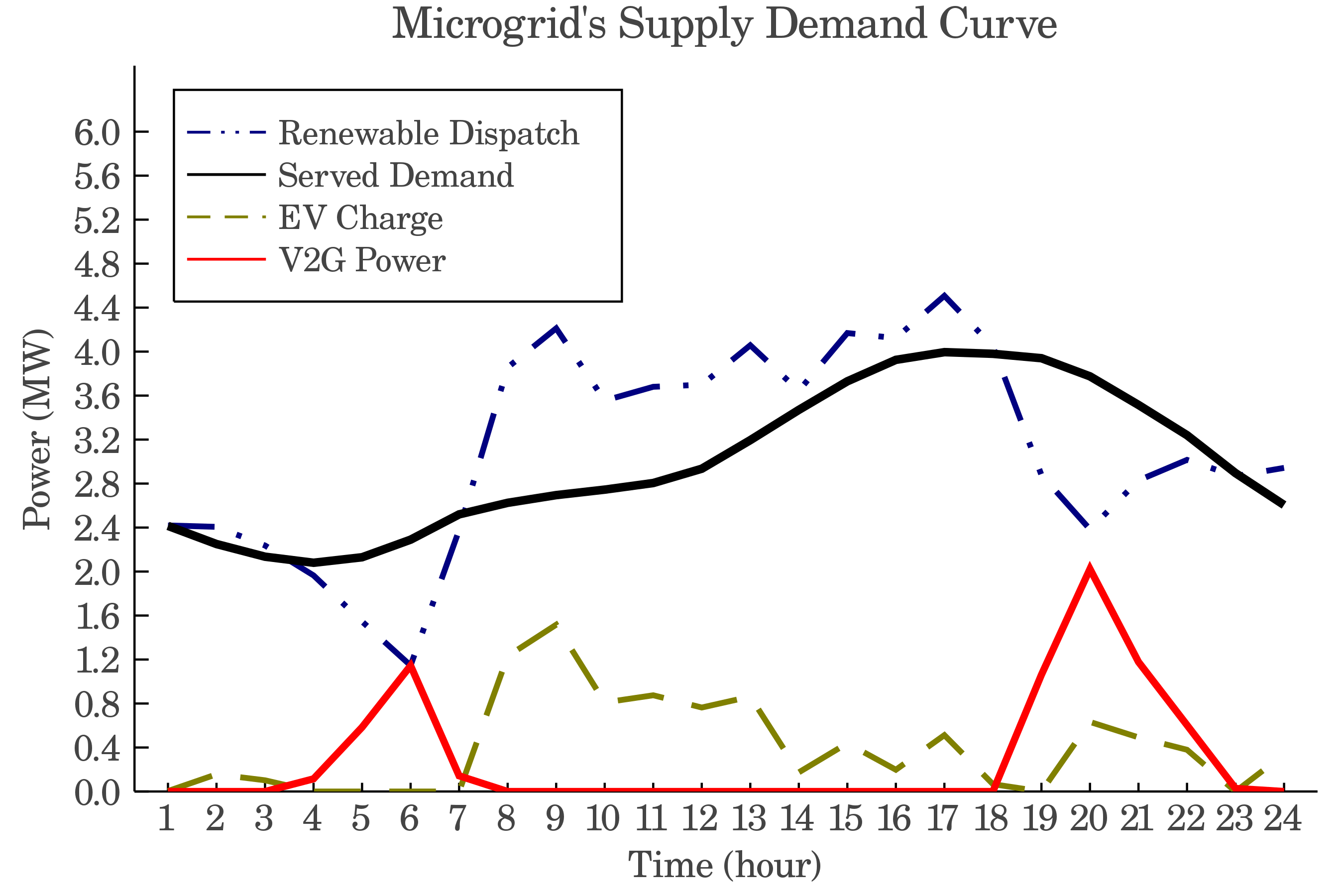}
        \vspace{-.35cm}
    \caption{Supply Demand Curve of Microgrid in Normal Conditions}
    \label{fig:case1_supply_demand}
    \vspace{.45cm}
\includegraphics[width=\linewidth]{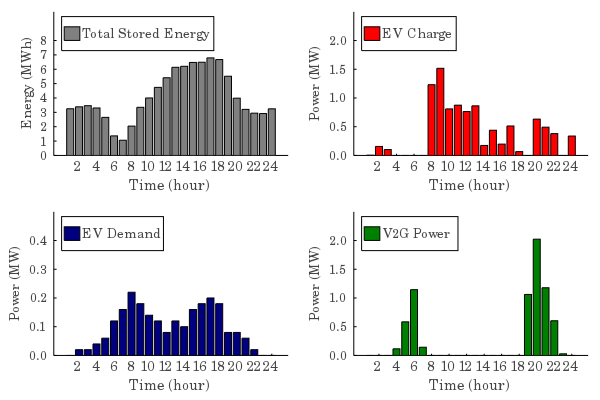}
         \vspace{-.85cm}
 \caption{Power and Energy of EVs in Normal Conditions}
  \label{fig:case1_ev_spec}
      \vspace{-.2cm}
\end{figure}

Results for normal daily operation problem of the proposed structure in presence of 480 EVs are displayed in Figures \ref{fig:case1_supply_demand} and \ref{fig:case1_ev_spec}. One important advantage of placing EV charging stations at DC side is making use of the fast charging technology. Latest EV technologies such as ``fast-charging" allow EVs to be charged/discharged with the rate of up to 250 kW \cite{collin2019advanced}. We consider the maximum power transmission rate of 150 kW between EV and microgrid in this work.\\
According to Fig. \ref{fig:case1_supply_demand}, the total output of renewable resources inside the microgrid falls below the costumer demand during hours 4-7 in the morning and 19-23 during the night. It is noticed that EVs can successfully be managed to discharge their excess stored energy during these hours and assist generation fleet in supporting the required demand. Here, all of the costumers' demand as well as EVs' demand are fully provided throughout the day. During sunny hours of the day (hours 8-18), PV units' output is increased and system's generation exceeds the system demand. The dashed line in Fig. \ref{fig:case1_supply_demand} corresponds to the energy charged by EVs, and is equal to the amount of surplus power provided by renewable suppliers during these hours. Fig. \ref{fig:case1_ev_spec} illustrates the total stored energy, energy required for daily trip, and charge and discharged amount of energy for all EVs in the system. One interesting observation in this plot is the coincidence of the overall EVs' charge and V2G power during hours 20-22 of the day. This is caused by operational constraints not allowing EVs at a certain location to be charged as much as they require during the day. Alternatively, EVs from other locations are charged and discharge their extra energy to other EVs later on.

\subsection{Case 2 - Analysis of V2G Degradation Cost}
\begin{figure}[h!]
    \centering
    \includegraphics[width=.82\linewidth]{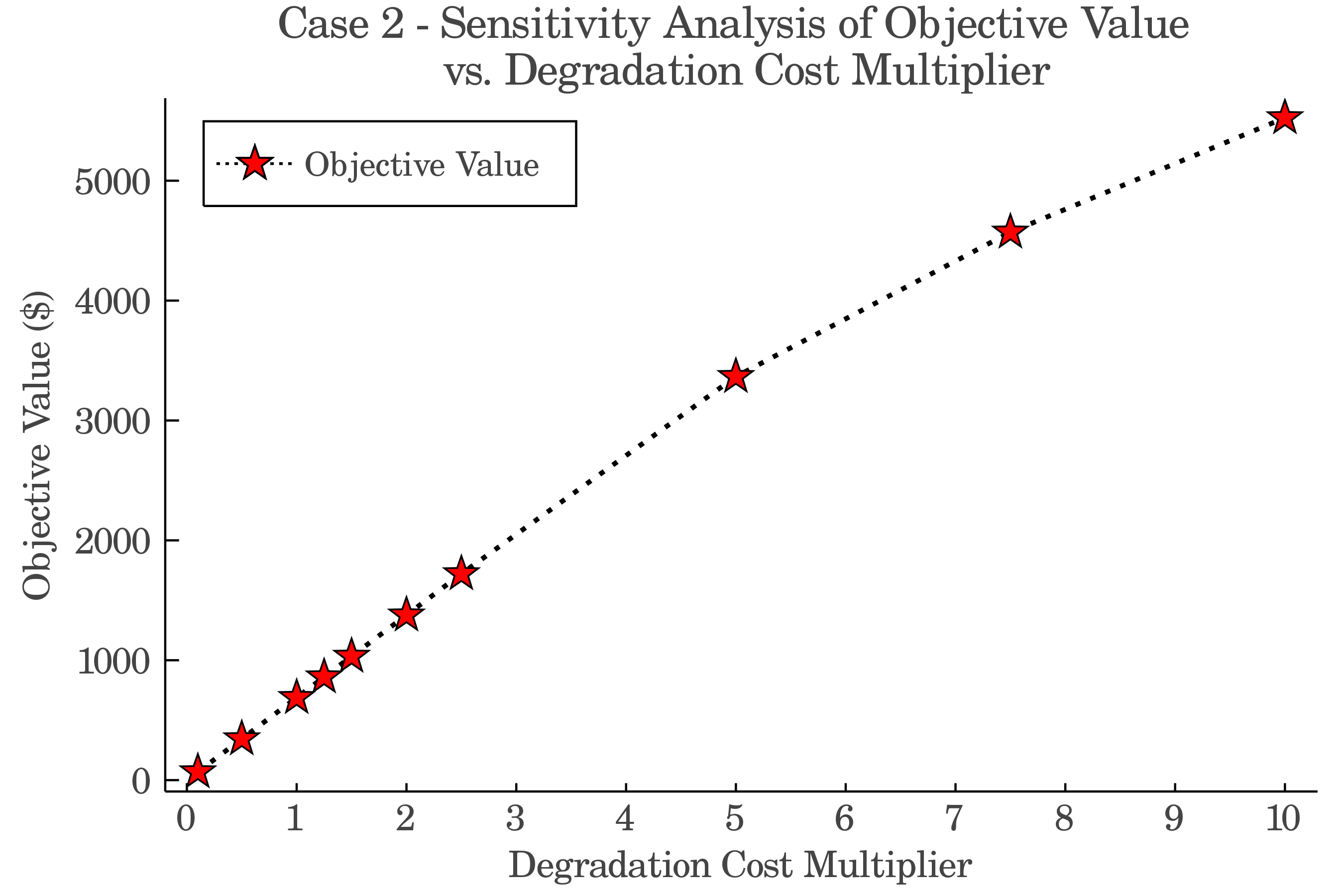}
         \vspace{-.35cm}
   \caption{Effect of Degradation Cost Multiplier on Objective Value}
    \label{fig:case2_deg_obj}
    \vspace{.65cm}
    \includegraphics[width=.82\linewidth]{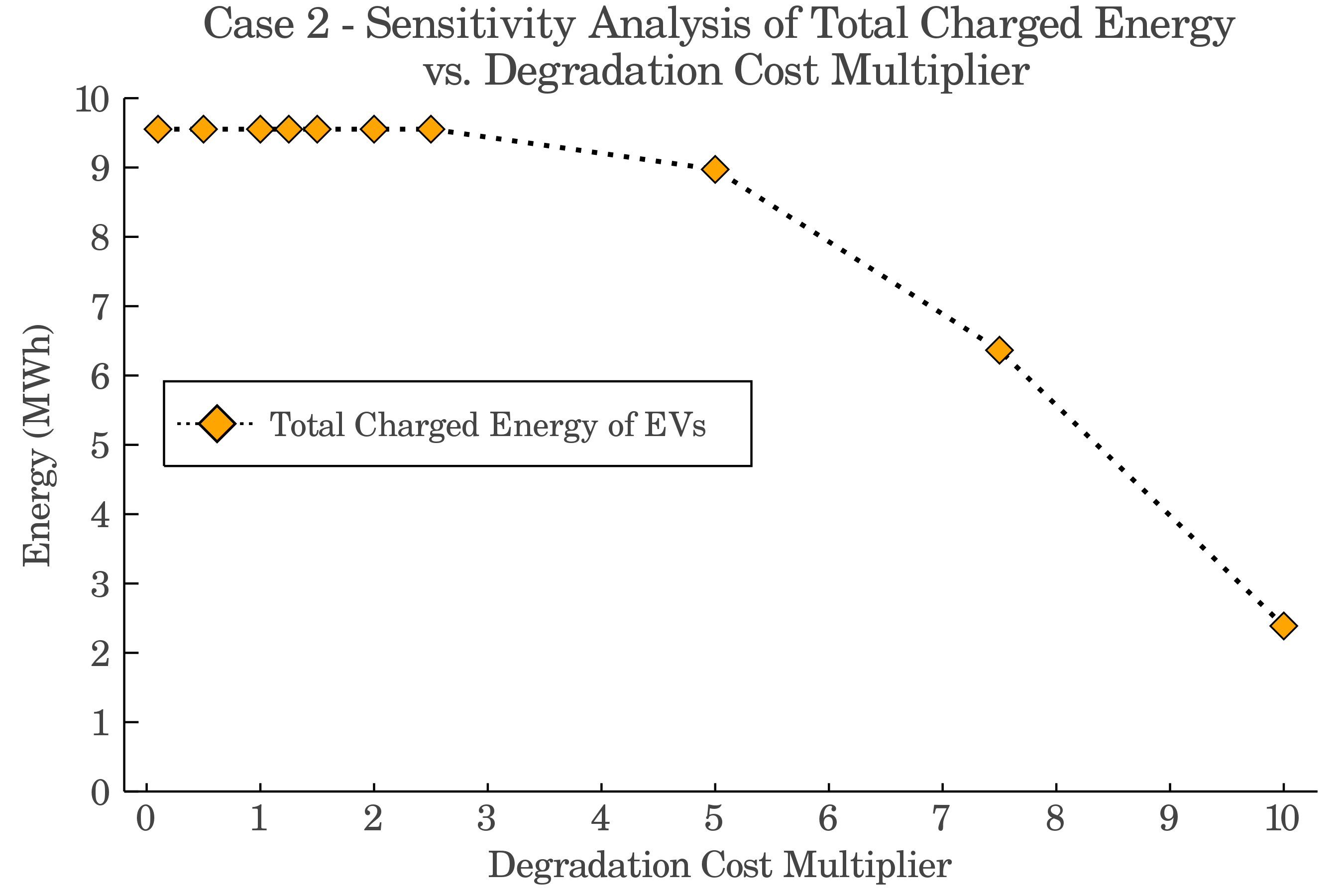}
         \vspace{-.35cm}
   \caption{Effect of Degradation Cost Multiplier on Charged Energy}
    \label{fig:case2_deg_charge}
    \vspace{.65cm}
    \includegraphics[width=.82\linewidth]{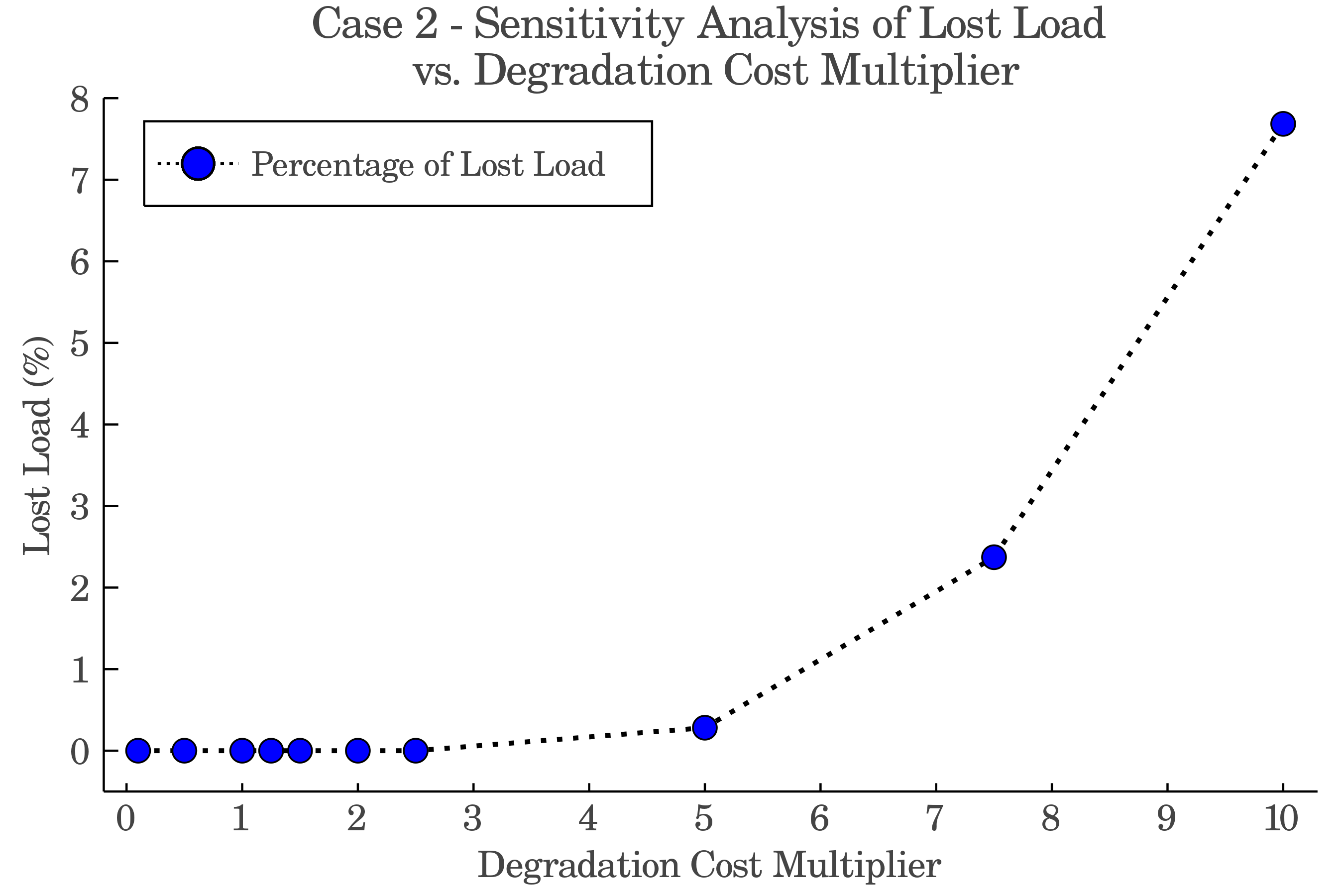}
         \vspace{-.35cm}
   \caption{Effect of Degradation Cost Multiplier on Lost Load}
    \label{fig:case2_deg_lost}
    \vspace{.65cm}
    \includegraphics[width=.82\linewidth]{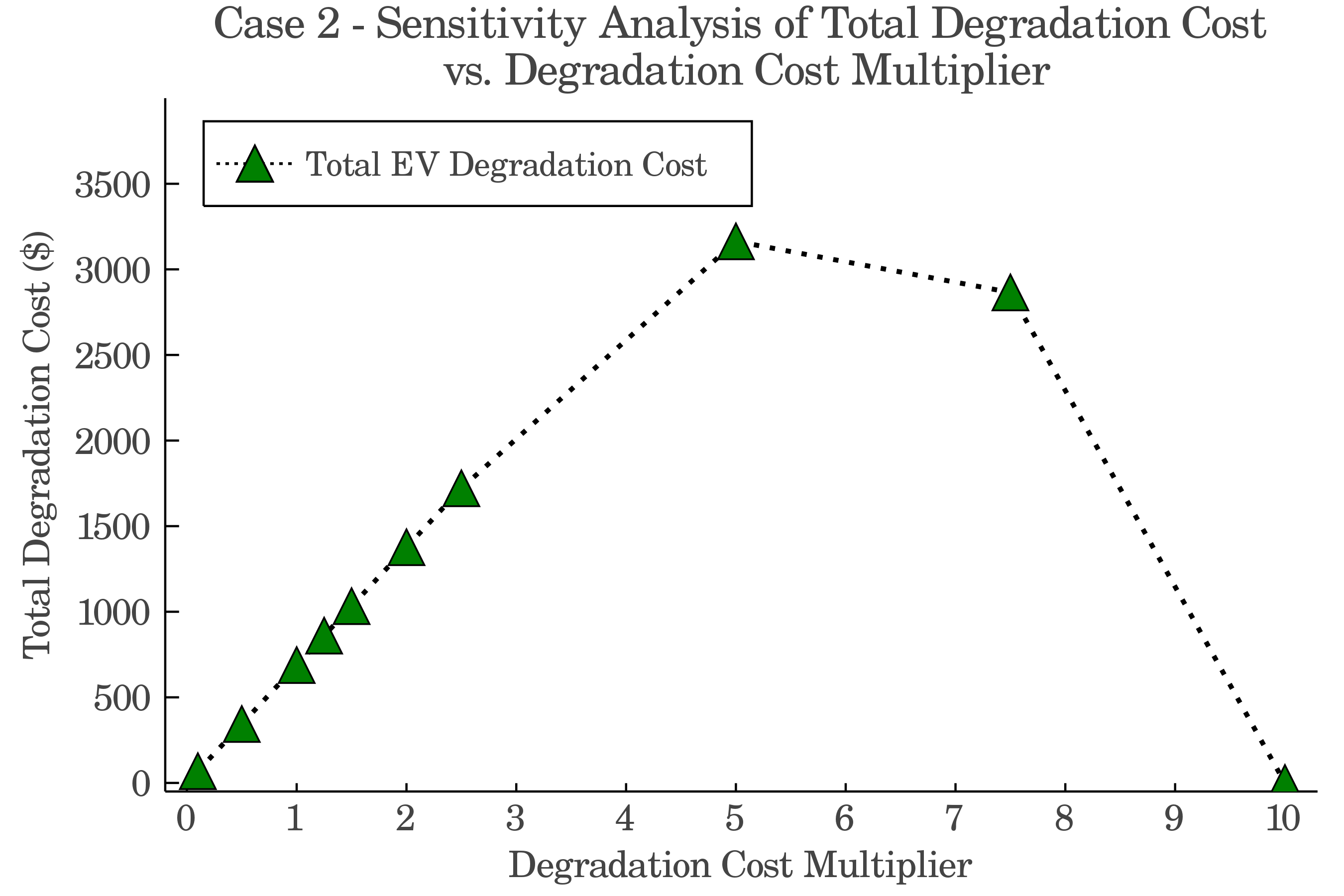}
         \vspace{-.35cm}
   \caption{Effect of Degradation Cost Multiplier on Total Degradation Cost}
    \label{fig:case2_deg_deg}
\end{figure}

In this case, the consequences of decreasing or increasing the rate of degradation cost are investigated for the daily operation of the microgrid. The results for this case are displayed in Figures \ref{fig:case2_deg_obj}-\ref{fig:case2_deg_deg}. In these observations, the degradation cost was multiplied by the factors in Table \ref{table:sens_coeff} for 10 instances. According to Fig. \ref{fig:case2_deg_obj}, the total objective value is to roughly proportional to the degradation cost multiplier. It is noticed that by either rise/fall in the degradation cost factor, the objective value also becomes lower/higher. This implies that the advancements in battery technology could lead to substantial savings by reduction in degradation costs. It is noteworthy to mention that since the objective value is composed of value of lost load and EV degradation costs, this figure in its own can not reveal information on behavior of these components.

\setlength\tabcolsep{2.2pt}
\begin{table}[t!]
\caption{Multiplier Coefficients for different case studies}
\resizebox{\linewidth}{!}{
\begin{tabular}{lcccccccccc}\hline\hline
Instance& 1 & 2 & 3& 4& 5 & 6& 7 & 8& 9 & 10\\ \hline
Case 2 - Degradation Cost& 0.1 & 0.5 & 1& 1.25 & 1.5 & 2& 2.5 & 5& 7.5 & 10\\
Case 3 - Solar Irradiance (\%) & 10& 20& 30 & 40 & 50& 60 & 70& 80 & 90& 100 \\
Case 4 - Inverter Capacity (\%) & 0& 15& 30 & 40 & 50& 60 & 70& 80 & 90& 100 \\\hline\hline
\label{table:sens_coeff}
\end{tabular}
}
\vspace{-.4cm}
\end{table}

In Figures \ref{fig:case2_deg_charge} and \ref{fig:case2_deg_lost}, the total charged energy by EVs and the percentage of lost load in case 2 are shown respectively. It is noticed that reduction in costs does not affect the charged power, as the system under normal scenario is already providing all the demand. By increasing the cost multiplier of degradation up to 2.5 times, it can be observed in both these figures that there is no change in the amount of energy charged in EVs and still all of the load is served. However, by increasing the degradation multiplier to 5 and more, the EV batteries' operation is affected and they start to charge less energy. At the same time, it is observed that at point 5, almost 0.3\% of the system's required load is not served. This situation is aggravated in cases with multiplier factor of 7.5 and 10, where the lost load percentage is 2.4\% and 7.7\%, respectively. Finally, the total degradation cost of EVs (second term in the objective function) in case 2 is displayed in Fig. \ref{fig:case2_deg_deg}. Here, it is observed that for the last instance, the total degradation cost is zero. This means EVs in this case only charge the amount of energy which is required for their daily trips and do not participate in any V2G service, as a result of extremely high degradation rates.

\subsection{Case 3 - Analysis of The Solar Irradiance Impact}
\begin{figure}[b!]
    \centering
    \includegraphics[width=.82\linewidth]{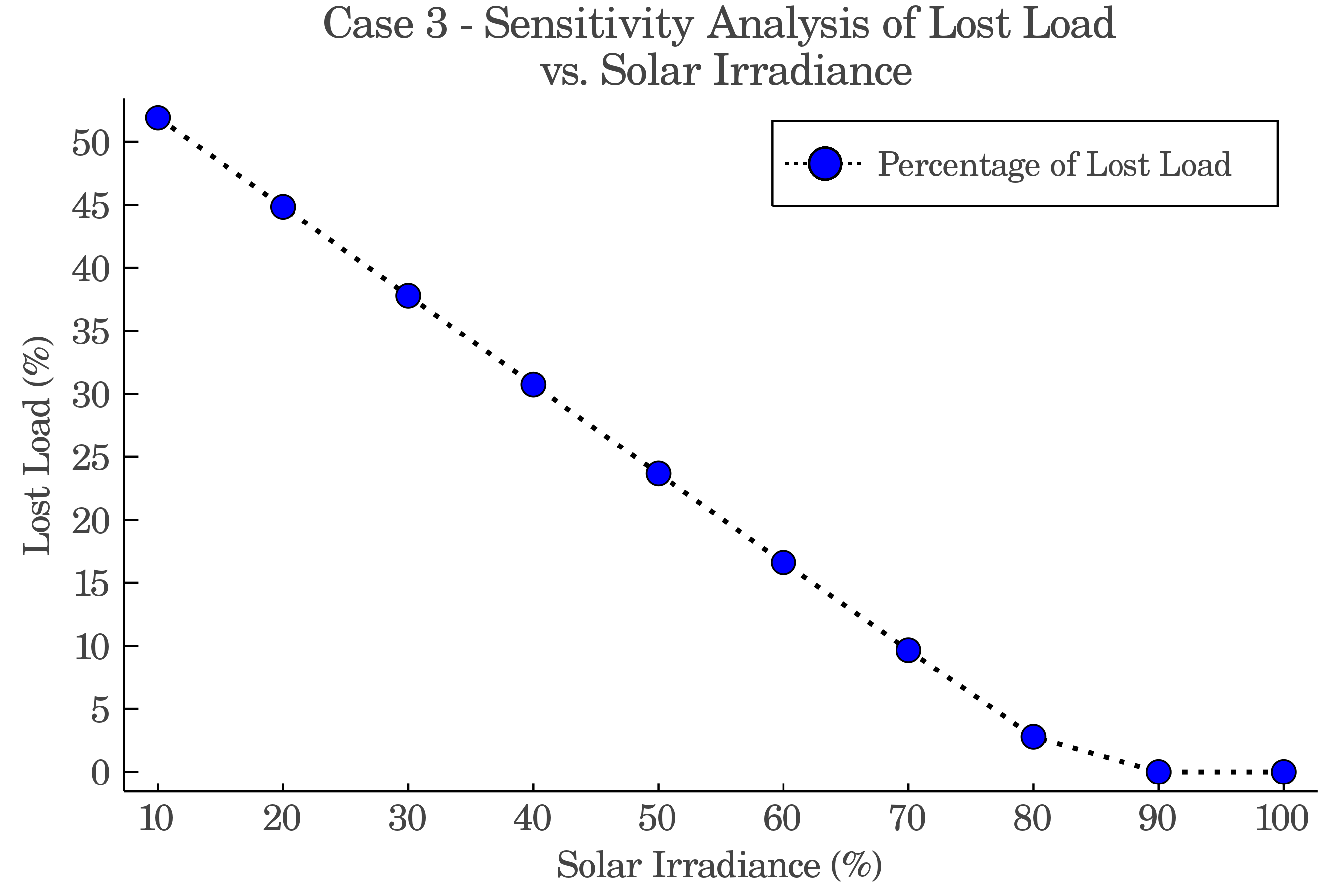}
         \vspace{-.35cm}
   \caption{Effect of Solar Irradiance on Lost Load}
    \label{fig:case3_pv_lost}
    \vspace{.65cm}
    \centering
   \includegraphics[width=.82\linewidth]{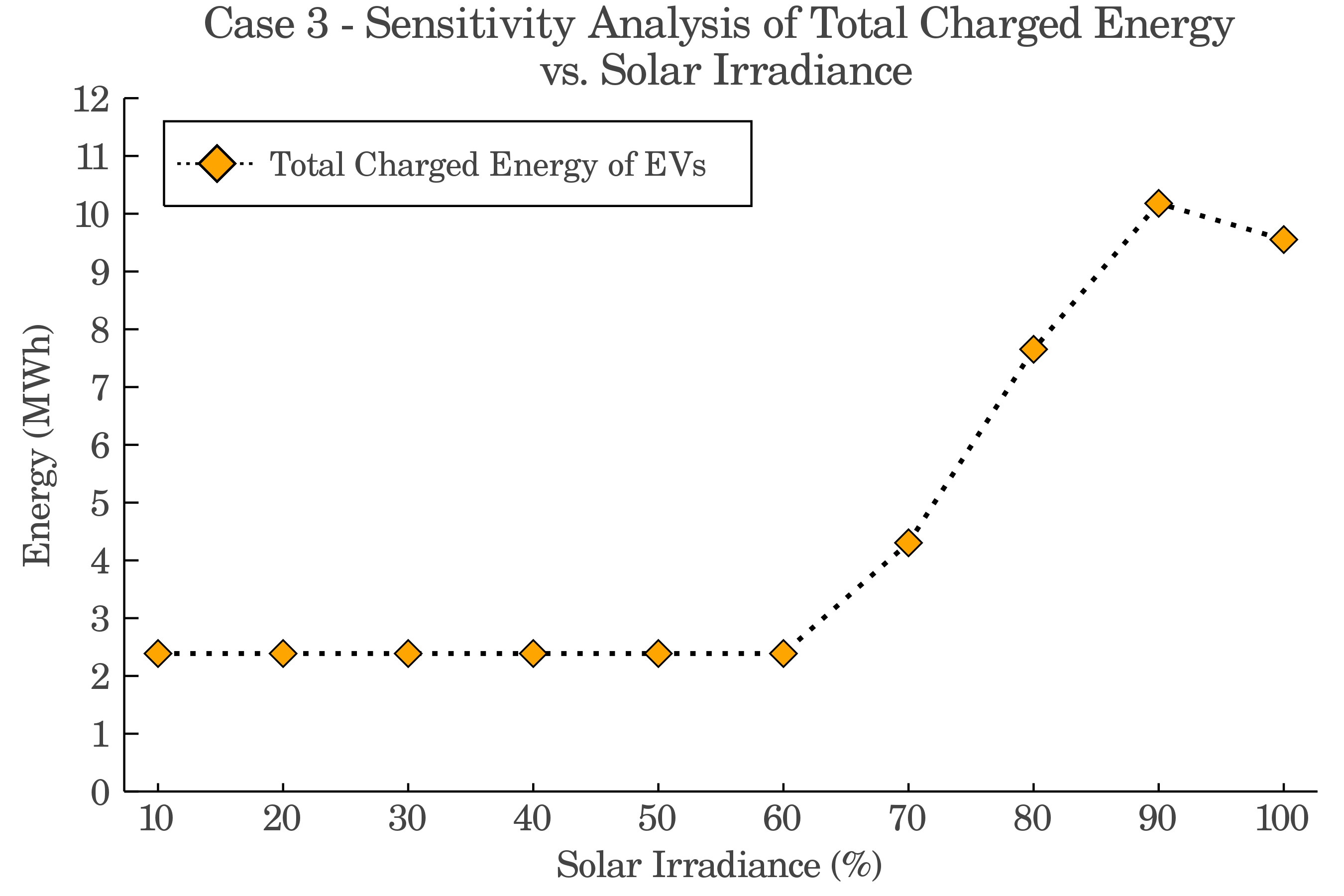}
         \vspace{-.35cm}
   \caption{Effect of Solar Irradiance on Charged Energy}
    \label{fig:case3_pv_charge}
\end{figure}

In this case, it is assumed that as a result of change in weather conditions, the PV output will reduce. Here, the daily microgrid operation under different levels of solar irradiance are considered, spanning from 10 to 100 percent (compared to the normal conditions). The results for this case are displayed in Figures \ref{fig:case3_pv_lost} and \ref{fig:case3_pv_charge}. Fig. \ref{fig:case3_pv_lost} displays the amount of lost load in case 3. According to this figure, by 10\% reduction in PV output, the microgrid is still capable of providing all of the required demand. However, further drops in solar irradiance and consequently PV outputs lead to considerable amounts of lost load, as the microgrid is highly dependant on PV generation. For example in case 1, almost 60\% of the microgrid's energy is provided by PV units.\\
In Fig. \ref{fig:case3_pv_charge}, total energy charged by EVs is displayed against different levels of solar irradiance. First of all, it is observed that for solar irradiance levels of 10-60\%, the charged energy by EVs is the same value. In fact, this is equal to the amount of energy that is required for daily commuting of EVs. It is only at 70\% solar irradiance and higher that EVs start to store excess energy for V2G purposes. Up until this point, no excess solar output was available for EVs to utilize their storage capabilities. It is interesting to notice that by dropping solar irradiance from 100\% to 90\%, the amount of energy charged by EVs is increased. In the case of 90\% irradiance, the network still serves all of the demand. This observation shows that the flexibility offered by EV storage is capable of mitigating mild reductions in renewable outputs. In the case of 90\% PV output, overall 4,490 MWh of solar energy is utilized, as opposed to the 4,430 MWh solar energy utilized in the case with 100\% PV output.

\subsection{Case 4 - Analysis of Inverters' Capacity}
\begin{figure}[b!]
    \centering    \includegraphics[width=.82\linewidth]{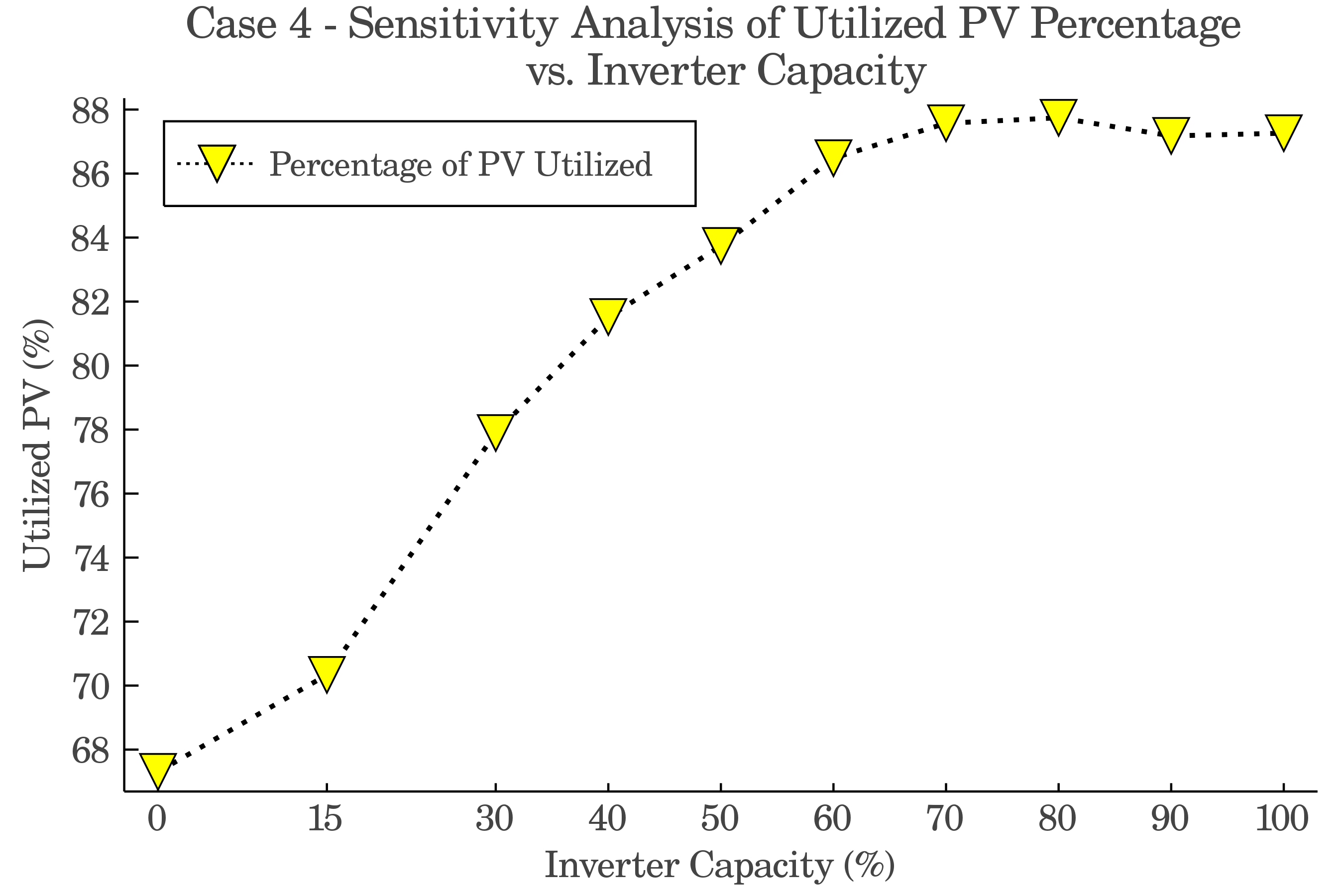}
         \vspace{-.35cm}
   \caption{Effect of Inverter's Capacity on PV Utilization}
    \label{fig:case4_inv_pv}
    \vspace{.65cm}
    \includegraphics[width=.82\linewidth]{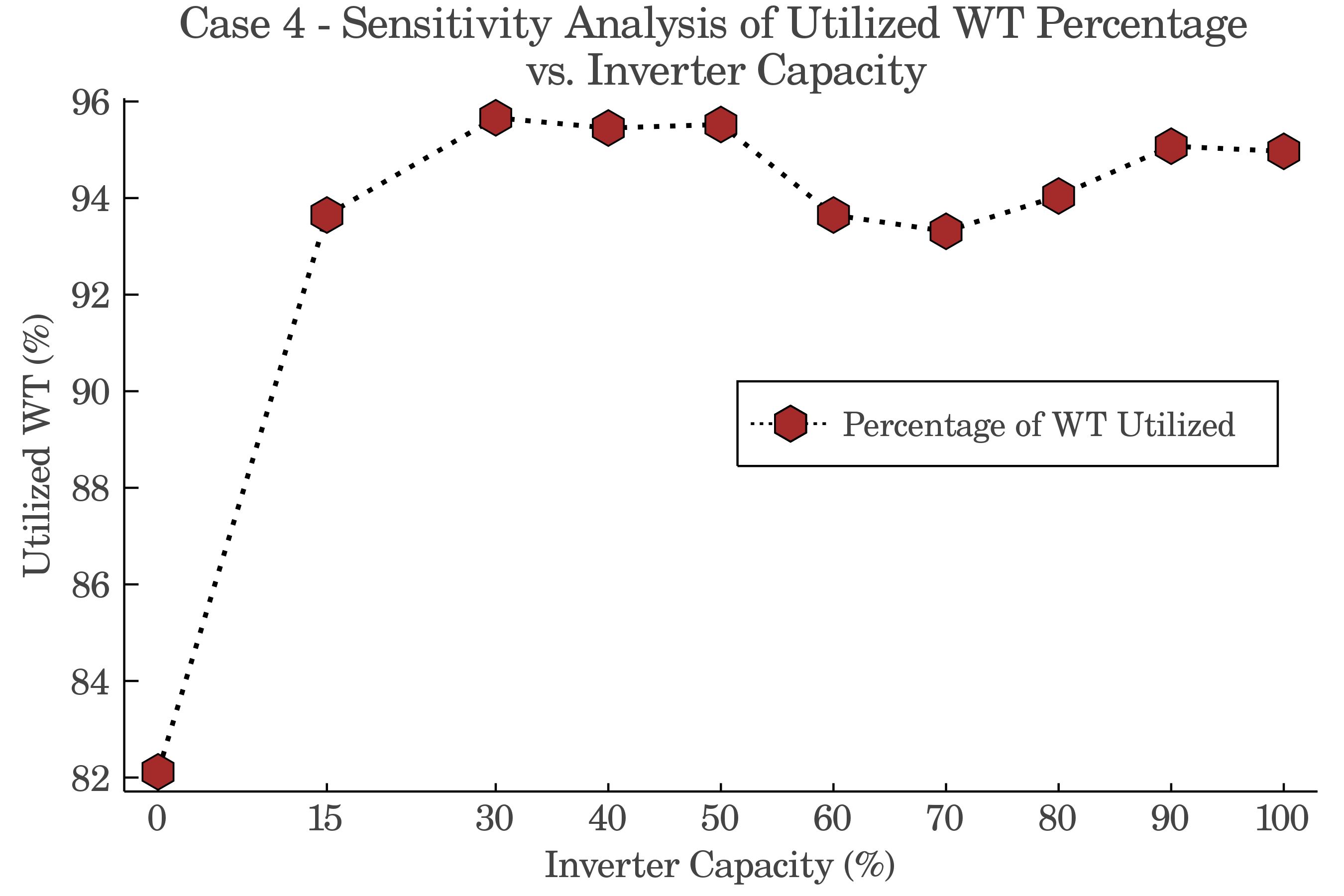}
         \vspace{-.35cm}
   \caption{Effect of Inverter's Capacity on WT Utilization}
    \label{fig:case4_inv_wt}
\end{figure}
In the proposed microgrid structure, the inverters connect AC and DC sides together and provide the means for reliable operation of the system. As a results of aging or failures, the amount of inverter capacity can be subject to reductions, which will negatively affect the microgrid operations. In case 4, the consequences of capacity reduction of inverters are explored for capacity values according to Table \ref{table:sens_coeff}. The amount of utilized PV and WT of system under changes in inverters' capacity are shown in Figures \ref{fig:case4_inv_pv} and \ref{fig:case4_inv_wt}, respectively. According to the structure of the microgrid, utilization of wind turbines is not affected by inverters. However, the utilization of PV units is correlated with the inverters' capacity. This is mostly due to the facts that first, the EV charging stations are placed in the DC side of the system. Also, most of battery charging takes place during the daylight hours, which means PVs are the main supplier of energy which is stored. Consequently, less exchange rate between AC and DC side means less energy can be stored in EVs and the renewable output should be used for matching the real-time load, rather than storage. But most of PV output happens during low-load conditions.\\

\begin{figure}[t!]
     \includegraphics[width=.82\linewidth]{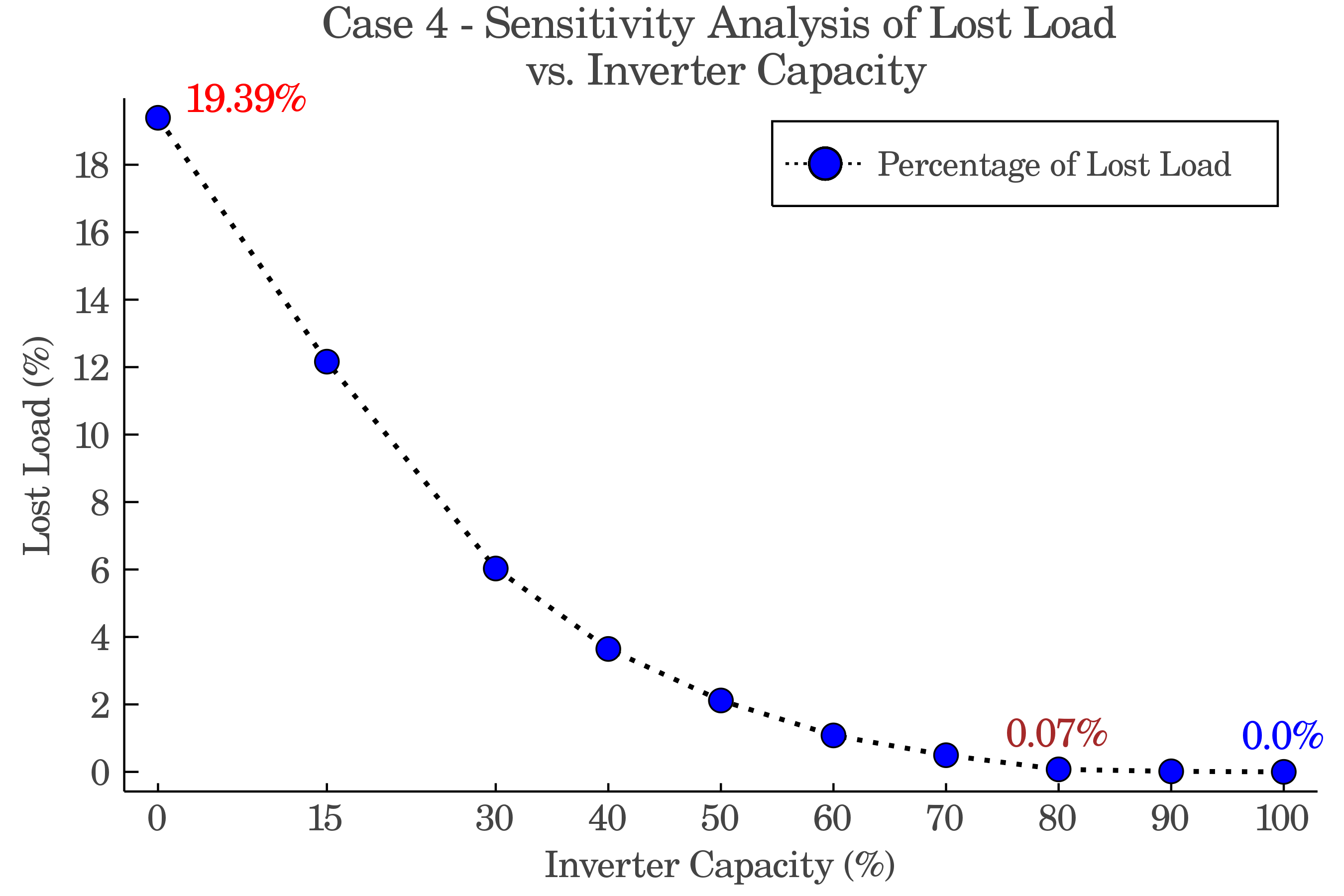}
         \vspace{-.35cm}
   \caption{Effect of Inverter's Capacity on Lost Load}
    \label{fig:case4_inv_lost}
    \vspace{.65cm}
    \includegraphics[width=.82\linewidth]{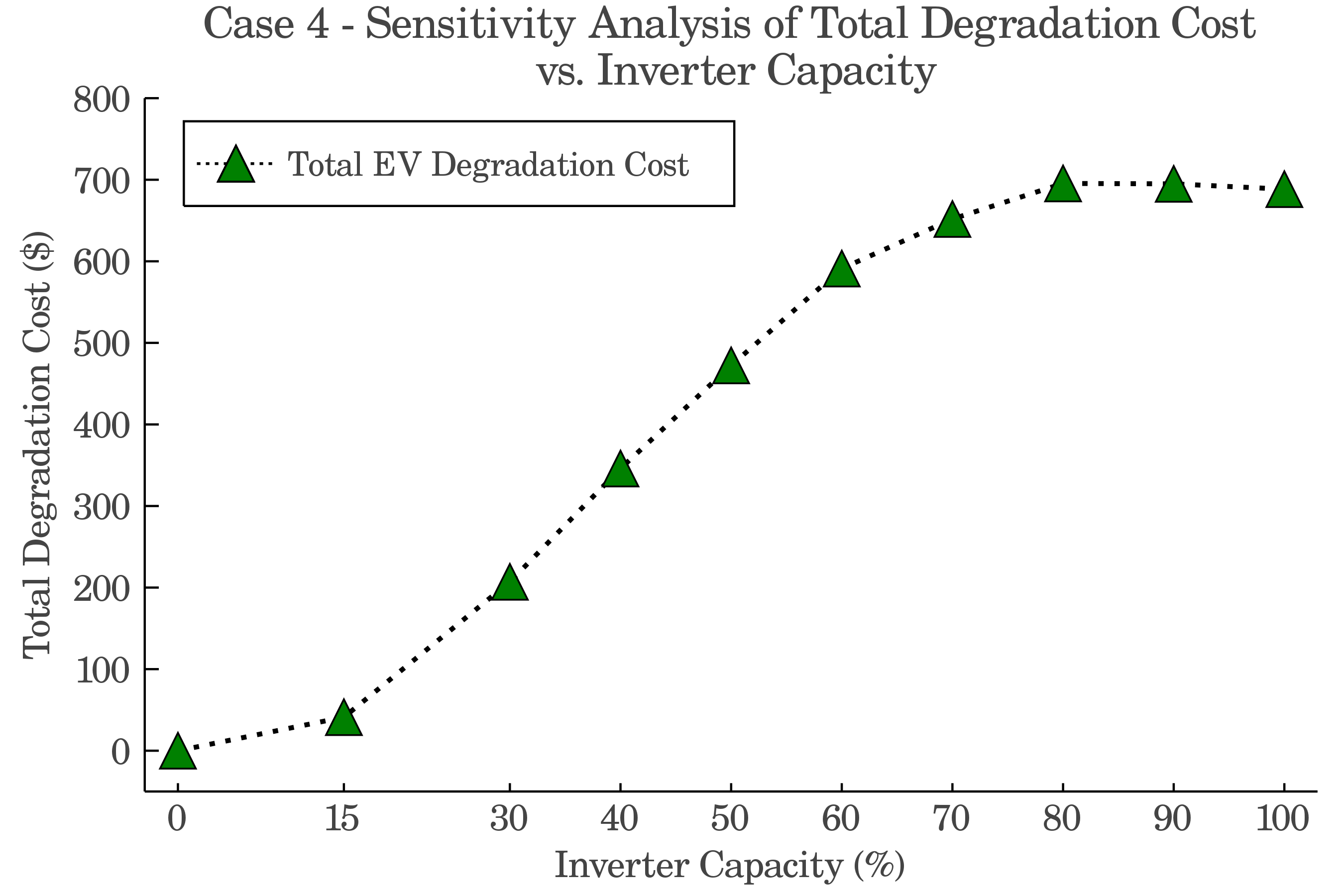}
         \vspace{-.35cm}
   \caption{Effect of Inverter's Capacity on Charged Energy}
    \label{fig:case4_inv_charge}
\end{figure}

Figures \ref{fig:case4_inv_lost} and \ref{fig:case4_inv_charge} show the amount of lost load and energy charged by EVs in daily operation of system under case 4 scenarios, respectively.
It is noticed from Fig. \ref{fig:case4_inv_lost} that with reduction in inverters' capacity, the amount of served load is also reduced. It is observed that when no energy is exchanged between AC and DC sides, 19.39\% of load is missed. According to Fig. \ref{fig:case4_inv_lost}, the V2G cost is also dependent on inverter capacity. With lower exchange possibility, lower energy is stored. The curve in Fig. \ref{fig:case4_inv_lost} displays a similar pattern to that of Fig. \ref{fig:case4_inv_pv}. This similarity indicates that PV utilization is closely related with the energy stored in batteries.

\section{Conclusion and Future Works}
In this work, a zero-carbon emitting hybrid AC/DC microgrid structure is investigated. Several charging lots are considered in the DC side of the microgrid to facilitate system efficiency. Storage systems are crucial for isolated microgrids as they store excess energy during excess output periods and inject it back to the system during insufficient output periods. It is shown that in a microgrid with no storage systems, EVs with fast-charging battery technologies can be utilized to reach optimal system performance. The case studies show that the proposed structure is capable of serving all of the required demand even in cases with lower PV outputs, higher V2G costs, and lower inverter capacity. This supports the fact that EV batteries are a means of offering flexibility for the reliable operation of the microgrid in real world situations. By accurate management of PEVs during their plugged-in times, they can satisfy energy for daily trips as well as microgrid's requirements. Applying scenario based algorithms for considering uncertainty in behavior of EV owners or renewable generation patterns could be a future work. Also, investing the planning problem for optimal charging station placement, and number and capacity of EVs is another possible extension to this research work. Finally, investigating problems that consider certainties in renewable generation and EV travel patterns can also be another path for future work.

\bibliographystyle{IEEEtran}

\bibliography{coord_sched_zc.bbl}

\begin{thebibliography}{10}
\providecommand{\url}[1]{#1}
\csname url@samestyle\endcsname
\providecommand{\newblock}{\relax}
\providecommand{\bibinfo}[2]{#2}
\providecommand{\BIBentrySTDinterwordspacing}{\spaceskip=0pt\relax}
\providecommand{\BIBentryALTinterwordstretchfactor}{4}
\providecommand{\BIBentryALTinterwordspacing}{\spaceskip=\fontdimen2\font plus
\BIBentryALTinterwordstretchfactor\fontdimen3\font minus
  \fontdimen4\font\relax}
\providecommand{\BIBforeignlanguage}[2]{{%
\expandafter\ifx\csname l@#1\endcsname\relax
\typeout{** WARNING: IEEEtran.bst: No hyphenation pattern has been}%
\typeout{** loaded for the language `#1'. Using the pattern for}%
\typeout{** the default language instead.}%
\else
\language=\csname l@#1\endcsname
\fi
#2}}
\providecommand{\BIBdecl}{\relax}
\BIBdecl

\bibitem{abazari2019coordination}
A.~Abazari, H.~Monsef, and B.~Wu, ``Coordination strategies of distributed
  energy resources including fess, deg, fc and wtg in load frequency control
  (lfc) scheme of hybrid isolated micro-grid,'' \emph{International Journal of
  Electrical Power \& Energy Systems}, vol. 109, pp. 535--547, 2019.

\bibitem{ghasemi2020multi}
A.~Ghasemi, A.~Shojaeighadikolaei, K.~Jones, M.~Hashemi, A.~G. Bardas, and
  R.~Ahmadi, ``A multi-agent deep reinforcement learning approach for a
  distributed energy marketplace in smart grids,'' in \emph{International
  Conference on Communications, Control, and Computing Technologies for Smart
  Grids (SmartGridComm)}.\hskip 1em plus 0.5em minus 0.4em\relax IEEE, 2020,
  pp. 1--6.

\bibitem{rahmanzadeh2018optimal}
M.~Rahmanzadeh, H.~Haggi, and M.~Aliakbar~Golkar, ``Optimal energy management
  of microgrid based on fcchp in the presence of electric and thermal loads
  considering energy storage systems,'' 2018.

\bibitem{shojaeighadikolaei2020demand}
A.~Shojaeighadikolaei, A.~Ghasemi, K.~R. Jones, A.~G. Bardas, M.~Hashemi, and
  R.~Ahmadi, ``Demand responsive dynamic pricing framework for prosumer
  dominated microgrids using multiagent reinforcement learning,'' \emph{North
  American Power Symposium (NAPS)}, pp. 1--6, 2020.

\bibitem{parhizi2015state}
S.~Parhizi, H.~Lotfi, A.~Khodaei, and S.~Bahramirad, ``State of the art in
  research on microgrids: A review,'' \emph{IEEE Access}, vol.~3, pp. 890--925,
  2015.

\bibitem{Khodayar2016TheArchitecture}
M.~Khodayar, S.~Manshadi, and A.~Vafamehr, ``{The short-term operation of
  microgrids in a transactive energy architecture},'' \emph{Electricity
  Journal}, vol.~29, no.~10, 2016.

\bibitem{manshadi2016decentralized}
S.~D. Manshadi and M.~Khodayar, ``Decentralized operation framework for hybrid
  ac/dc microgrid,'' in \emph{2016 North American Power Symposium
  (NAPS)}.\hskip 1em plus 0.5em minus 0.4em\relax IEEE, 2016, pp. 1--6.

\bibitem{bayani2020short}
R.~Bayani, M.~Bushlaibi, and S.~D. Manshadi, ``Short-term operational planning
  problem of the multiple-energy carrier hybrid ac/dc microgrids,'' \emph{2021
  IEEE PES General Meeting}, 2021.

\bibitem{bayani2020autonomous}
R.~Bayani, S.~D. Manshadi, G.~Liu, Y.~Wang, and R.~Dai, ``Autonomous charging
  of electric vehicle fleets to enhance renewable generation dispatchability,''
  \emph{CSEE Journal of Power and Energy Systems}, 2021.

\bibitem{ahmadian2020review}
A.~Ahmadian, B.~Mohammadi-Ivatloo, and A.~Elkamel, ``A review on plug-in
  electric vehicles: Introduction, current status, and load modeling
  techniques,'' \emph{Journal of Modern Power Systems and Clean Energy},
  vol.~8, no.~3, pp. 412--425, 2020.

\bibitem{babaei2021data}
M.~Babaei, A.~Abazari, M.~M. Soleymani, M.~Ghafouri, S.~Muyeen, and M.~T.
  Beheshti, ``A data-mining based optimal demand response program for smart
  home with energy storages and electric vehicles,'' \emph{Journal of Energy
  Storage}, vol.~36, 2021.

\bibitem{lekvan2021robust}
A.~A. Lekvan, R.~Habibifar, M.~Moradi, M.~Khoshjahan, S.~Nojavan, and
  K.~Jermsittiparsert, ``Robust optimization of renewable-based multi-energy
  micro-grid integrated with flexible energy conversion and storage devices,''
  \emph{Sustainable Cities and Society}, vol.~64, p. 102532, 2021.

\bibitem{ansari2019framework}
M.~Ansari, M.~Ansari, and A.~Asrari, ``A framework for simultaneous management
  of greenhouse gas emission and substation transformer congestion via
  cooperative microgrids,'' in \emph{2019 North American Power Symposium
  (NAPS)}.\hskip 1em plus 0.5em minus 0.4em\relax IEEE, 2019, pp. 1--6.

\bibitem{gurobi}
\BIBentryALTinterwordspacing
L.~Gurobi~Optimization, ``Gurobi optimizer reference manual,'' 2021. [Online].
  Available: \url{http://www.gurobi.com}
\BIBentrySTDinterwordspacing

\bibitem{collin2019advanced}
R.~Collin, Y.~Miao, A.~Yokochi, P.~Enjeti, and A.~von Jouanne, ``Advanced
  electric vehicle fast-charging technologies,'' \emph{Energies}, vol.~12,
  no.~10, p. 1839, 2019.

\end{thebibliography}

\end{document}